\documentclass[conference]{IEEEtran}

\usepackage[dvips]{color}
\usepackage{epsf}
\usepackage{times}
\usepackage{epsfig}
\usepackage{graphicx}

\usepackage{amsmath}
\usepackage{amssymb}
\usepackage{amsxtra}
\usepackage{amsthm}

\usepackage{here}
\usepackage{rawfonts}
\usepackage{times}
\usepackage{url}
\usepackage{cite}
\usepackage{comment}
\usepackage[utf8]{inputenc}
\usepackage{caption}
\usepackage{subcaption}

\usepackage{pstricks}
\usepackage{algorithm}
\usepackage{algpseudocode}

\usepackage
[
a4paper, 
left=1.59cm,
right=1.58cm,
top=2.1cm, 
]
{geometry}

\newcommand*\dif{\mathop{}\!\mathrm{d}}

\makeatletter

\begin{document}
 \title{Machine Learning for Predictive On-Demand Deployment of UAVs for Wireless Communications}\vspace{-0.1cm}	
\author{\IEEEauthorblockN{Qianqian Zhang$^1$, Mohammad Mozaffari$^1$, Walid Saad$^1$, Mehdi Bennis$^2$, and M\'erouane Debbah$^{3,4}$}
	
	\IEEEauthorblockA{\small
		$^1$Bradley Department of Electrical and Computer Engineering, Virginia Tech, VA, USA,
		Emails: \url{{qqz93,mmozaff,walids}@vt.edu}.\\
		$^2$Center for Wireless Communications, University of Oulu, Finland, Email: \url{bennis@ee.oulu.fi}.\\
		$^3$Mathematical and Algorithmic Sciences Lab, Huawei France R\&D, Paris, France, Email: \url{merouane.debbah@huawei.com}.\\
		$^4$Large Systems and Networks Group, CentraleSupélec, Université Paris-Saclay, 3 rue Joliot-Curie, 91192 Gif-sur-Yvette, France.  
		\vspace{-0.3cm}		
	}
}
\maketitle

\begin{abstract}
In this paper,  a novel machine learning (ML) framework is proposed for enabling a predictive,  efficient deployment of unmanned aerial vehicles (UAVs), acting as aerial base stations (BSs), to provide on-demand wireless service to  cellular users. 
In order to have a comprehensive analysis of cellular traffic, an ML framework based on a Gaussian mixture model (GMM) and a weighted expectation maximization (WEM) algorithm is introduced to predict the potential network congestion. 
Then, the optimal deployment of UAVs is studied to minimize the transmit power needed to satisfy the communication demand of users in the downlink, while also minimizing the power needed for UAV mobility, based on the predicted cellular traffic. 
To this end,  first, the optimal partition of service areas of each UAV is derived, based on a fairness principle.  
Next, the optimal location of each UAV that  minimizes the total power consumption is derived. 
Simulation results show that the proposed ML approach can reduce the required downlink transmit power and improve the power efficiency by over $20\%$, compared with an optimal deployment of UAVs with no ML prediction. 
\end{abstract}

\IEEEpeerreviewmaketitle

\section{Introduction}
The demand for cellular data is experiencing an unprecedented increase. 
The next generation, $5$G wireless cellular network is estimated to support a $200$ fold increase in wireless data traffic by 2030 \cite{rangan2014millimeter}. 
To cope with this exponential increase in demand, 
there has been growing interest in network densification for cellular systems as a means to improve spectrum efficiency and cellular network capacity  \cite{bhushan2014network}. 

The need for additional base stations (BSs) is more pronounced in cellular hotspot areas that exhibit a steep surge in data demands during temporary events, such as concerts and football games.  
To satisfy such temporary surges in traffic, the use of an unmanned aerial vehicle (UAV) as an aerial BS  can be a more flexible and cost-effective approach, compared with a traditional, ground BS \cite{mozaffari2018tutorial}. 
A mobile UAV can intelligently change its position, which is suitable to provide on-demand wireless service to ground users, thus overcoming 
coverage holes and alleviating congestions \cite{mozaffari2016optimal}.

In order to deploy UAVs in a timely and flexible manner, network operators must be able to predict potential hotspots and congestion events a priori. 
To this end, there is a need to apply  machine learning (ML) techniques to analyze  demand patterns \cite{chen2017machine}.  
The ability of ML to exploit big data analytics  enables a comprehensive prediction of a network's traffic amount and data distribution.  
By using such predictions,  aerial UAV BSs can be optimally deployed to the target area beforehand thus providing an on-demand, delay-free and power-efficient wireless service to ground users.

The use of  UAVs as cellular BSs has been addressed in \cite{mozaffari2016efficient, mozaffari2016optimal, becvar2017performance, li2016energy, sharma2016uav, zeng2017energy}.  
Meanwhile, in \cite{mozaffari2016optimal} and \cite{becvar2017performance}, the authors studied the use of UAVs as flying BSs to provide energy-efficient service to wireless users. Moreover, the work in \cite{li2016energy} and \cite{sharma2016uav} focus on using UAVs as relays, and the work in \cite{zeng2017energy} studies an energy-efficient trajectory design. 
However, most of the existing works assume a time-invariant wireless network, or a given distribution of cellular users.  
To properly  analyze an on-demand deployment of UAV BSs, the temporal and spatial patterns of the cellular traffic data must be predicted so as to optimally deploy  UAVs  to satisfy a time-varying data demand.

There are existing woks, such as \cite{sharma2016uav,horn2012neural}, and \cite{chen2017learning}, that apply ML techniques to optimize UAV deployment. 
In \cite{sharma2016uav}, a neural model is formulated to study the map of UAVs to each hotspot areas. 
The authors in \cite{horn2012neural} studied the trajectory optimization using neural networks, while a segmented regression approach is proposed in \cite{chen2017learning} for UAV channel modeling, based on the terrain topology. 
However, none of these works demonstrates the benefit of applying ML to deploy UAVs on-demand and improve power efficiency and network performance.  
In order to analyze the data traffic of cellular networks, the authors in \cite{zhang2017traffic} studied a BS sleeping strategy for minimizing power consumption.  However, the authors focused only on a low-traffic cellular network, which is not scalable for the more practical, congested scenarios.   

The main contribution of this paper is a novel machine learning framework that enables operators to predict congestions and hotspot events, and subsequently, deploy temporary UAV BSs to provide aerial wireless service to mobile users, while minimizing the UAV power needed for downlink communications and mobility. 
We consider a heterogeneous cellular network, in which ground BSs can offload the wireless service to aerial UAVs when the predicted data demand of mobile users exceeds the network capacity. 
To guarantee a no-delay wireless service, a Gaussian mixture model (GMM)  is introduced based on a weighted expectation maximization (WEM)  algorithm \cite{gebru2016algorithms}  to predict the cellular data traffic. 
Then, the optimal deployment of UAVs is studied to minimize the power needed for UAV transmission and mobility, given the predicted traffic. 
To this end, we first study the division of service areas, based on a fairness principle.  
Then, we derive the optimal UAV locations that can minimize the total power consumption of the network. 
To the best of our knowledge, this is the first work that leverages ML to predictively deploy UAVs  as aerial  BSs. 
Simulation results show that the proposed approach can reduce the required downlink transmit power and improve the power efficiency by over $20\%$, compared with an optimal deployment of UAVs with no ML prediction.

The rest of this paper is organized as follows. Section \ref{sysModel_proFormulation} presents the system model and problem formulation. Section \ref{solution} outlines the proposed ML and UAV deployment framework. Simulation results are presented in Section \ref{simulation}, while conclusions are drawn in Section \ref{conclusion}. 

\section{System Model and Problem Formulation}\label{sysModel_proFormulation}
Consider a time-variant heterogeneous cellular network that serves a group of cellular users distributed in a geographical area $\mathcal{A}$.  
The cellular network consists of a set $\mathcal{I}$ of $I$ UAVs and a set $\mathcal{J}$ of $J$ BSs. 
Each user can receive data from both ground BSs and UAVs. 
Initially, a traditional BS will be chosen to serve the wireless users. 
However, if the downlink of the ground cellular system is overloaded due to heavy traffic, the ground BS will request the deployment of UAV BSs to offload some of its users. 

Ground BSs and UAVs employ different frequency bands for downlink communications. 
Each UAV is equipped with directional antennas that enable beamforming. 
Therefore,  interference among  UAVs is negligible. 
Furthermore, each UAV adopts a frequency division multiple access (FDMA) technique and assigns a dedicated channel to one of its downlink users.  
Hereinafter, we use the notion of an \emph{aerial cell} to indicate the service area of each UAV, and \emph{aerial cellular users} to indicate users that are served by UAV  cellular BSs.  

Each UAV has a limited energy resource, that must be efficiently used for joint communications and mobility. 
To this end, the UAVs should intelligently change their positions to meet the required users' data rates, as well as to minimize their transmission power. 
However, given the cellular network is time-variant, the cellular traffic demand will change over time, which complicates the efficient deployment. 
To guarantee timely aerial service without having UAVs continuously moving, the network operator can use ML techniques to predict its network's data demand, and then, request the deployment of UAV BSs to the predicted hotspot areas,  before the congestion occurs. 

\subsection{Air-to-ground channel model}
Given a typical ground receiver located at $(x,y) \in \mathcal{A}$ and a UAV $i \in \mathcal{I}$ located at $(x_i,y_i,h_i)$, the path loss of the downlink communication from UAV $i$ to the receiver will be \cite{al2014optimal}: 
\begin{align}\label{pl}
L_i(x,y)[dB] = 20 \log \left( \frac{4 \pi f_c d_{i}(x,y)}{c} \right) + \xi _i,
\end{align}
where  $d_i(x,y) = \sqrt{(x-x_i)^2+(y-y_i)^2+h_i^2}$ is the distance between the ground receiver and UAV $i$, $f_c$ is the carrier frequency,  $c$ is the speed of light, and $\xi_i$ is the average additional loss to the free space propagation loss which depends on the environment.  
If the wireless link between UAV $i$ and a ground user is line-of-sight (LOS), $\xi^{\text{LOS}} _i \sim N(\mu_{\text{LOS}},\sigma^2_{\text{LOS}})$; otherwise, the non-line-of-sight (NLOS) link has an additional loss of $\xi^{\text{NLOS}} _i \sim N(\mu_{\text{NLOS}},\sigma^2_{\text{NLOS}})$. The NLOS link will experience a high path loss due to  shadowing  and reflection. 
The probability of existence of LOS links between UAV $i$ and the ground user will then be \cite{al2014optimal}:
\begin{align}\label{probLOS}
p^{\text{LOS}}_i(x,y) = \frac{1}{1+ a \exp \left( -b[ \frac{180 }{\pi} \theta_i(x,y)-a  ]\right)},
\end{align}
where $a$ and $b$ are constant values which depend on the environment, and $\theta_i(x,y) =\sin ^{-1} (\frac{h_i}{d_i(x,y)}) $ is the elevation angle of UAV $i$ with respect to the receiver. Then, the probability of having a NLOS link is $p^{\text{NLOS}}_i(x,y) = 1-p^{\text{LOS}}_i(x,y)$ \cite{mozaffari2016optimal}. 

Consequently, the average path loss from UAV $i$ to the ground reciever at $(x,y)$ in the linear scale can be given as
\begin{align}
\bar{L}_i(x,y) = p^{\text{LOS}}_i(x,y)  L^{\text{LOS}}_i(x,y)  + p^{\text{NLOS}}_i(x,y)  L^{\text{NLOS}}_i(x,y).
\end{align}
Therefore, the downlink capacity that UAV $i$ can provide to a mobile user located at $(x,y)$ will be:
\begin{align}\label{capacity}
R_i(x,y) = W_i \log_2\left( 1+ \frac{P_i(x,y) G_i(x,y)/ \bar{L}_i(x,y)}{W_in_0} \right),
\end{align}
where $W_i$ is the transmission bandwidth of UAV $i$, $P_i(x,y)$ is the transmission power, $G_i(x,y)$ is the antenna gain of UAV $i$, and $n_o$ is the average noise power spectral density. 
For tractability, we assume a perfect beam alignment between the UAV and the mobile receiver, and each UAV has the same antenna gain. Therefore, $G_i(x,y) = G$, which is a constant for all $ i \in \mathcal{I}$ and $(x,y) \in \mathcal{A}$. 
Assume that the total available bandwidth of UAV $i$ is $B_i$ and the number of mobile users associated with UAV $i$ is $N_i$, then the downlink bandwidth of each channel will be $W_i = B_i/N_i$.

The number  $N_i$ of aerial users that are served by UAV $i$ within its aerial cell is given by: 
\begin{align}
N_i =  N \int \int_{\mathcal{A}_i} f(x,y) \dif x \dif y,
\end{align}
where $N = \sum_{i \in \mathcal{I}}N_i$ is the total number of aerial users, $\mathcal{A}_i$ is the   service area of UAV $i$, and $f(x,y)$ is the distribution of aerial users.
In order to provide a universal wireless service, the aerial cells of all UAVs should fully cover the geographical area $\mathcal{A}$ without overlap. 
That is, $\cup_{i \in \mathcal{I}} \mathcal{A}_i = \mathcal{A}$, and for $i \ne j$ and $i,j \in \mathcal{I}$, $\mathcal{A}_i \cap \mathcal{A}_j = \emptyset$. 
However, note that, the value of $N$ and the user distribution $f(x,y)$ will change, according to the offloading requests from ground BSs. 

\subsection{Cellular traffic analysis} \label{cellularTrafficSec}

To provide an on-demand service,  network operators need to change the UAVs' locations frequently, according to the offload requests from ground BSs, to satisfy the instant traffic demand.    
However, such continuous movement will consume excessive power.  
To efficiently deploy UAVs while guaranteeing  a no-delay wireless service, a dataset of the cellular traffic history can be exploited by the network operator for traffic prediction. 
This dataset, represented by a matrix $\boldsymbol{Q}$, records discrete data during each time period $T$ for $M$ days:  
\begin{align}
\boldsymbol{Q} =  [N(x,y,t),~ D(x,y,t)|\forall t \in \mathcal{T}, (x,y) \in \mathcal{A} ],
\end{align}
where $\mathcal{T} = \{ T, 2T, \cdots, 24M\}$ is a discrete set of time, and the unit of $T$ is hour.  
The first item $N(x,y,t)$ represents the number of aerial users that are offloaded from a BS at $(x,y)$ to a UAV during a time interval from $t$ to $t+T$, and the second item $D(x,y,t)$ denotes the amount of cellular traffic that a UAV needs to provide for the aerial users from a BS at $(x,y)$ during the period from $t$ to $t+T$.  

Let $N$ be the total number of aerial users, $D$ be the total amount of aerial cellular traffic, $f(x,y)$ be the spatial distributions of aerial users,  and $g(x,y)$  be the spatial distribution of aerial data traffic in $\mathcal{A}$.  
Without a comprehensive analysis of $\boldsymbol{Q}$, the values of $N$, $D$, $f(x,y)$ and $g(x,y)$ will change over time, based on the offloading requests of ground BSs, which causes a frequent movement of UAVs to meet the instant traffic demand, and excessive power consumed on mobility. 

Therefore, our goal is to develop a centralized ML approach to predict $N$ and $f(x,y)$  based on $N(x,y,t)$, and  $D$ and $g(x,y)$ based on $D(x,y,t)$, such that at the beginning of each period $T$, network operators can optimally deploy UAVs to minimize the power consumptions, while during each interval the locations of UAVs remain fixed.  

\subsection{Data rate requirement}  
Given  the predicted information on the total amount of aerial cellular traffic $D$, and the distribution of aerial cellular traffic $g(x,y)$, the average data rate  requirement within a service area $\mathcal{A}_i$ of UAV $i$ can be given by
\begin{align}\label{demand}
\alpha_i = \frac{1}{T} \int \int_{\mathcal{A}_i} D \cdot g(x,y) \dif x \dif y.
\end{align}  
Since the communication capacity of UAV $i$ should be greater than or equal to the rate demand of all users in its aerial cell $\mathcal{A}_i$, we formulate the data rate requirement as follows, 
\begin{align}
\int \int_{\mathcal{A}_i} R_i(x,y) \dif x \dif y   \ge \alpha_i, 
\end{align}
i.e.,
\begin{align} \label{rateRequire}\vspace{-0.1cm}
R_i(x,y)  \ge \frac{Dg(x,y)}{T}.  
\end{align}
We define $\beta(x,y) =  \frac{Dg(x,y)}{T}$ as the average minimum data rate requirement for the aerial user at $(x,y)$. 
Based on (\ref{capacity}) and (\ref{rateRequire}), the minimum transmit power that UAV $i$ should provide to communicate with the user at $(x,y)$ will be: 
\begin{align}\label{minPower}
P^{\text{min}}_i (x,y)= \frac{B_i n_0 \bar{L}_i(x,y) }{GN_i }  \left( 2^{\beta(x,y)N_i/B_i} -1 \right). 
\end{align}
Note that, the values of $N_i$ and $\beta(x,y)$ in (\ref{minPower}) will depend on the output of the cellular traffic analysis.  

Consequently, the total transmit power of all UAVs needed to satisfy the data demand of all aerial users in $\mathcal{A}$ will be: 
\begin{align}
P_c & = \sum_{i \in \mathcal{I}} \int \int_{\mathcal{A}_i}    P^{\text{min}}_i (x,y)    \dif x \dif y . 
\end{align}
Without loss of generality, we assume that the maximum transmission power of UAVs is sufficient to meet the data demand of aerial users. 
Meanwhile, the total power for each UAV $i \in \mathcal{I}$ to move from its current location $(x^o_i,y^o_i,h^o_i)$ to the new location $(x_i,y_i,h_i)$ will be: 
\begin{align}
	P_t = \gamma  \sum_{i \in \mathcal{I}} \left[ (x^o_i - x_i)^2 + (y^o_i - y_i)^2 +(h^o_i-h_i)^2 \right] ^{\frac{1}{2}},
\end{align}
where  $\gamma$ is the rate of energy consumption a UAV needs to move by one meter.

Then, the second objective is to jointly find the optimal location and the partition of the service area $\mathcal{A}_i$ for each UAV $i \in \mathcal{I}$, such that  the total power used for downlink transmissions and mobility can be minimized, i.e.,
\begin{subequations}\label{powerMin}
\begin{align} 
\min_{\mathcal{A}_i,x_i,y_i,h_i} \quad & P_c + P_t,\\
\textrm{s. t.} \quad & \frac{\int \int_{\mathcal{A}_i}    P^{\text{min}}_i (x,y)    \dif x \dif y}{P^{\text{a}}_i} = \kappa, \forall i \in \mathcal{I}, \\
 & \cup_{i \in \mathcal{I}} \mathcal{A}_i = \mathcal{A}, \\
 & \mathcal{A}_i \cap \mathcal{A}_j = \emptyset, \forall i \ne j \in \mathcal{I},
\end{align}
\end{subequations}
where $P^{\text{a}}_i$ is the available power of UAV $i$, and $\kappa$ is a constant for all $i \in \mathcal{I}$. The first constraint represents a fairness principle, whereby the ratio of the data traffic offloaded to each UAV equals to the ratio of the available power of each UAV. 
The second and third constraints guarantee that the service areas of all UAVs fully cover $\mathcal{A}$ without overlap. 

Note that, without an ML analysis, the function $P_i^{\text{min}}$, as well as $P_c$, will change, based on the offloading requests of  ground BSs.  
Thus, the network operator needs to reorganize the aerial cellular system to meet the instant traffic demand frequently.   
However, with the predicted information of cellular traffic, the optimal problem (\ref{powerMin}) is  fixed within each period $T$. Therefore, at the beginning of each interval, UAVs are deployed according to the solution of (\ref{powerMin}), and within the period, the location and aerial cell of each UAV remain unchanged. 
 
\section{Proposed Prediction and UAV Deployment Framework} \label{solution}
Next, we propose a novel approach to address the aforementioned problems.
First, a centralized  ML approach will be proposed to predict the values of $N$, $D$, $f(x,y)$ and $g(x,y)$ for each time interval $T$. With the prediction information, the power minimization problem in (\ref{powerMin}) will be solved to optimally deploy each UAV.  

\subsection{Cellular traffic prediction} \label{learn}
 
In order to have a robust and practical analysis, we use the real dataset \footnote{Our approach can accommodate other datasets without loss of generality.} of City Cellular Traffic Map \cite{cityCellularTrafficMap}, which records 
the time, the location of each BS,  the number of mobile users, and the total amount of data that  each BS serves during each hour, from Aug. 19 to Aug. 26, 2012,  in a median-size city in China.  
We assume that the maximum number of mobile users that each BS can serve within one hour is a fixed number of $N_m$, and the maximum amount of cellular data is a constant $D_m$ for all BSs. 
Thus, a new dataset is generated to capture the traffic of the aerial cellular network as  
$\boldsymbol{Q}^{'} = [N(x,y,t)-N_m,~ D(x,y,t)-D_m|\forall t \in \mathcal{T}, (x,y) \in \mathcal{A} ]$, 
in which $N(x,y,t)-N_m$ is the number of aerial users from hour $t$ to $t+1$, and $ D(x,y,t)-D_m$ is the amount of aerial cellular traffic.  
For notation simplicity, hereinafter, we use  $\boldsymbol{Q}$ to denote the aerial traffic dataset, instead of $\boldsymbol{Q}^{'}$. 
Since $N(x,y,t)$ and $D(x,y,t)$ have an analogous data structure, a similar approach will be applied to analyze $N(x,y,t)$ and $D(x,y,t)$. For simplicity, we keep the following discussion only on $D(x,y,t)$. 
Therefore, the objective is to use ML to formulate the temporal and spatial pattern of $D(x,y,t)$.

There are three key assumptions in the following ML analysis.  
First, due to the periodicity of human activity, the cellular traffic  presents a repetitive daily pattern \cite{paul2011understanding}.  
Based on this observation, we assume that the total cellular traffic during a specific hour of different days follows the same distribution. 
Therefore, we divide the dataset  into $24$ subsets, by merging  
the data  of the same hour from different days.  
Second, we assume that the traffic amount between each hour of one day is independent. Therefore, given the $24$ sub-datasets, $24$ independent models will be built to study the pattern of each objective value of each hour.  
Furthermore, we assume that the temporal feature of $D(x,y,t)$ is independent from the spatial distribution. 
As a result, two separate models will be studied to identify the temporal feature $D(t)$ and the spatial feature $g(x,y)$ of $D(x,y,t)$ for each hour.

The model  to capture the temporal and spatial characteristics of $D(x,y,t)$ relies on a  GMM, which assumes that the data distribution can be modeled by the sum of multiple Gaussians with different weights  as \cite{christopher2016pattern} $ p(x) = \sum^{K}_{k=1} \pi_k \mathcal{N}(x| \mu_k,\Sigma_k)$, 
where $x$ is a general data point, $p(x)$ is the probability distributed at $x$,   
$K$ is the number of individual Gaussian models in GMM, and  $\sum_{k=1}^{K} \pi_k= 1$, $\pi_k \in [0,1]$ is the  mixing coefficient for each Gaussian.

\vspace{0cm}
\begin{algorithm}[t] \footnotesize
	\caption{Proposed algorithm to find parameters of the spatial distribution model $g^t(x,y)$} \label{algo2}
	\begin{algorithmic}
		\State \textbf{Input}:  $\boldsymbol{G}^t$ for a given $t$ 
		\State \textbf{Output}: $\{\pi_k\}, \{ \boldsymbol{\mu}_k\}$ and $\{\boldsymbol{ \Sigma}_k\}$, for each $k \in \{1,\cdots ,K\}$\\
		\\
		\textbf{Part I}:  Weighted K-means for parameter initialization 
		\State \textbf{Input}: $\boldsymbol{G}^t$ 
		\State \textbf{Output}: $K$, $\{ \boldsymbol{\mu}_k\}_{k \in \{1,2,\cdots,K\}}$\\
		A. \textbf{For} $K = K_\text{min}:K_\text{max}$\\
		\quad 1. Randomly choose $K$ initial values of $\{\boldsymbol{ \mu}_k\}_{k \in \{1,\cdots,K\}}$, \\
		\quad 2. \textbf{Loop}:\\
		\quad \quad a. Calculate the weighted distance of each data point to each $\boldsymbol{\mu}_k$ by \\
		\quad \quad \quad  $D(x,y,t)|\boldsymbol{x}-\boldsymbol{\mu}_k|$,\\
		\quad \quad b. Assign each data point $\boldsymbol{x}$ to cluster $k^*$, such that \\ 
		\quad \quad \quad $k* = \arg \min_k D(x,y,t)|\boldsymbol{x}-\boldsymbol{\mu}_k|$, \\
		\quad \quad c. Recalculate  $\boldsymbol{\mu}_k$ by averaging  the values of data points belonging \\
		\quad \quad \quad  to cluster $k$ as $\boldsymbol{\mu}_k = \sum_{\mathcal{C}_k} D(x,y,t) \boldsymbol{x}  / \sum_{\mathcal{C}_k} D(x,y,t) $,\\
		\quad \quad d. Check convergence: if $\{ \boldsymbol{\mu}_k\}_{k \in \{1,2,\cdots,K\}}$ changes.\\
		B. Choose the value of $K$ that minimizes the ratio of intra-cluster to inter-\\
		\quad cluster distance \cite{christopher2016pattern}.
		\\
		\\
		\textbf{Part II}: Weighted  EM iteration 
		\State \textbf{Input}: $\boldsymbol{G}^t$,  $K$, $\{ \boldsymbol{\mu}_k\}_{k \in \{1,2,\cdots,K\}}$
		\State \textbf{Output}: $\{\pi_k\}, \{ \boldsymbol{\mu}_k\}$ and $\{\boldsymbol{ \Sigma}_k\}$, for each $k \in \{1,\cdots ,K\}$\\
		1. \textbf{Initialize} $\boldsymbol{\Sigma}_k$ to be an identical matrix, and $\pi_k = 1/K$. \\
		2. \textbf{E step}: Calculate the posterior probability for each data   point $\boldsymbol{x}_n$ \\
		\quad belonging to each cluster $k$ by \\
		\quad $r_{nk} = \pi_k \mathcal{N}(\boldsymbol{x}_n|\boldsymbol{\mu}_k,\boldsymbol{\Sigma}_k) / \sum_{i=1}^{K}\pi_i \mathcal{N}(\boldsymbol{x}_n|\boldsymbol{\mu}_i,\boldsymbol{\Sigma}_i)$,\\
		3.\textbf{ M step}: Recalculate the parameters using the posterior probability $r_{nk}$ \\
		\quad \quad  a. $\boldsymbol{\mu}_k = \sum_{n=1}^{N}  D(x,y,t) r_{nk} \boldsymbol{x}_n /N_k$\\
		\quad \quad  b. $\boldsymbol{\Sigma}_k = \sum_{n=1}^{N} D(x,y,t) r_{nk} ( \boldsymbol{x}_n - \boldsymbol{\mu}_k)( \boldsymbol{x}_n - \boldsymbol{\mu}_k)^T/N_k$\\
		\quad \quad  c. $\pi_k = N_k/K$\\
		\quad where $N_k = \sum_{n=1}^{N}D(x,y,t) r_{nk}$.\\
		4. Check the \textbf{convergence} by (\ref{loglike}). If not converged, return to E step.  
	\end{algorithmic}
\end{algorithm}
\subsubsection{Spatial distribution model}\label{MLg}
First, we study the modeling approach of  the spatial feature $g(x,y)$ of $D(x,y,t)$.
Given a time $t \in \{1,2,\cdots, 24\}$, the data distribution of the cellular traffic from $t$ to $t+1$ can be calculated by 
\begin{align} \vspace{0cm}
g^t(x,y) = \frac{ D(x,y,t)}{ \int \int_{\mathcal{A}} D(x,y,t) \dif x \dif y}.
\end{align} 
Then, a dataset $\boldsymbol{G}^t$ is formed by all the distribution $g^t(x,y)$ of $M$ days for the specific hour $t$, and we seek to build a GMM to capture a pattern of data distribution for time $t$ as 
\begin{align}
g^t(\boldsymbol{x}) =  \sum^{K^t}_{k=1} \pi^t_k  \mathcal{N}(\boldsymbol{x}| \boldsymbol{ \mu}^t_k,\boldsymbol{\Sigma^t}_k),
\end{align} 
where $\boldsymbol{x} = (x,y)$ is the location vector.
To find the parameters of  $K^t$, $\pi^t_k$, $\boldsymbol{\mu}^t_k$, and $\boldsymbol{ \Sigma}^t_k$, for a given $t \in \{1,2,\cdots,24\}$, and $k \in \{1,\cdots, K^t\}$,  
first, a classification approach based on a weighted K-means method is used to group the data $\boldsymbol{x}$ into $K$ clusters, and the weight $D(x,y,t)$ is the data amount at $\boldsymbol{x} = (x,y)$.  
Then, the WEM algorithm will be used to find the optimal parameters of GMM. 
The convergence of the WEM iterative approach can be evaluated by the log likelihood function  as 
\begin{align} \label{loglike}
\ln \mathcal{L} (\boldsymbol{ \Sigma},\boldsymbol{ \mu}, \pi) = \sum_{n} \ln \sum_k \pi_k D(\boldsymbol{x}_n)g^t(\boldsymbol{x}_n|\boldsymbol{ \Sigma}_k,\boldsymbol{ \mu}_k),  
\end{align}
whose value will increase as the iteration time increases. 
Our detailed approach is summarized in Algorithm \ref{algo2}.

\subsubsection{Temporal distribution model}\label{MLD}
Given a time $t$, the total aerial traffic amount in the system from $t$ to $t+1$ can be calculated by $D^t = \sum_{(x,y) \in \mathcal{A}} D(x,y,t)$. 
By gathering the data $D^t$ of $M$ days, we have a dataset $\boldsymbol{D}^t = \{D^t_1,\cdots,D^t_M\}$. 
The GMM that captures the temporal pattern of $\boldsymbol{D}^t$ is $p(D^t) = \sum^{V^t}_{v=1} \pi^t_v \mathcal{N}(D^t| \mu^t_v,\Sigma^t_v)$.   
The approach to model the temporal distribution $D^t$ for $D(x,y,t)$,  is similar to the algorithm in Algorithm \ref{algo2}. However, both the K-means and EM algorithm do not add weight to each data point. 
As a result, by ignoring all $D(x,y,t)$  used in Table \ref{algo2} and substituting its value by one, Algorithm \ref{algo2} can be applied to find the temporal pattern $D^t$. 
The mixture Gaussian model $p(D^t)$  is a probability density function (pdf) over the cellular data amount, from which we can get the cumulative distribution function (CDF) as $C_t(d) = \int_{-\infty}^{d} p(D^t) \dif D^t$. 
The predicted data amount can be estimated by the CDF with a threshold. For example, with a threshold of $60\%$, the predicted traffic amount over the aerial networks  can be given by $D^t = C_t^{-1}(0.6)$.   
The ML analysis of the temporal feature $N(t)$ and the spatial feature $f(x,y)$ of $N(x,y,t)$ can follow the approach of Algorithm \ref{algo2}.

\subsection{On-demand, optimal UAV deployment}

In order to optimally deploy UAVs to minimize the total power, problem (\ref{powerMin}) is formulated, which jointly considers the aerial cell partition and the UAVs' locations. 
With the prediction information, network operators only need to move UAVs at the beginning of each time interval, according to the solution of (\ref{powerMin}). 
However, solving (\ref{powerMin}) is challenging due to the mutual dependence between $(x_i,y_i,h_i)$ and $\mathcal{A}_i$ with $N_i$ and $\beta(x,y)$.   
For tractability, we solve (\ref{powerMin}) in two sequential steps. 
First, given the current location of each UAV $i \in \mathcal{I}$, we seek to find the optimal partition of the service area $\mathcal{A}_i$ for each UAV, that minimizes the power for transmissions. 
Then, for each UAV $i$, given its fixed service area $\mathcal{A}_i$, the optimal location is derived to minimize the required power for downlink communications and mobility. 

\subsubsection{Optimal partition of service areas}  
Given the current location of each UAV $i \in \mathcal{I}$, we aim to find the best partition of service areas $\{\mathcal{A}_i\}_{i\in \mathcal{I}}$, such that the total power for downlink communications of all UAVs is minimized. 
The optimal partition problem can be formulated as follows,
\begin{subequations}\label{powerMinPartition}
\begin{align}
& \underset{\mathcal{A}_i}{\text{min}} 
& &   P_c, \\
& \text{s. t.} & & \int \int_{\mathcal{A}_i}    P^{\text{min}}_i (x,y)    \dif x \dif y   = \kappa P^{\text{a}}_i, \forall i \in \mathcal{I}, \\
& & & \cup_{i \in \mathcal{I}} \mathcal{A}_i = \mathcal{A}, \\
& & & \mathcal{A}_i \cap \mathcal{A}_j = \emptyset, \forall i \ne j \in \mathcal{I}.
\end{align}
\end{subequations}
To solve this problem, we use our previously developed gradient-based method in \cite[Theorem 1, Algorithm 1]{mozaffari2017wireless}.

\subsubsection{Optimal locations}
Given the optimal partition of the service area $\{\mathcal{A}_i\}_{i \in \mathcal{I}}$, the power minimization problem can be reduced into $I$ subproblems for each UAV $i \in \mathcal{I}$ as 
\begin{equation}\label{powerMinEachTwo}
\begin{aligned}
& \underset{x_i,y_i,h_i}{\text{min}} 
& &   P_i^c +  P_i^t .
\end{aligned}
\end{equation}
Based on \cite[Theorem 1]{mozaffari2016optimal}, we  focus on two scenarios in the following discussions. One is a high-altitude UAV, where $h_i^2 \gg (x-x_i)^2 + (y-y_i)^2$, and the other is the low-altitude UAV, where $h_i^2 \ll (x-x_i)^2 + (y-y_i)^2$. 
In scenario one, the value of $\theta_i$ in (\ref{probLOS}) is approximately $90^{\circ}$, thus, $p^{\text{LOS}}_i(x,y) \approx 1$ and $\bar{L}_i(x,y) \approx L^{\text{LOS}}_i(x,y)$. 
Then, $P_i$ can be rewritten as
\begin{equation}
\begin{aligned}
P^c_i  \approx  O_i \int \int_{\mathcal{A}_i}   Z_i(x,y)  \left( d^2_{i}(x,y)  + 10^{0.1\xi^{\text{LOS}} _i} \right)   \dif x \dif y,
\end{aligned}
\end{equation}
where $O_i = \left( \frac{4 \pi f_c}{c}\right)^2 \frac{B_i n_0  }{GN_i } $ is a coefficient that does not depend on $(x,y)$, and $Z_i(x,y) = 2^{\beta(x,y)N_i/B_i} -1$. 
It is obvious that $P_i$ is a convex function with respect to $x_i$ and $y_i$. 
By setting the first partial derivatives to be zero, we have the optimal locations for UAV $i$ that  minimize the transmission power $P_i^c$ as 
\begin{subequations}\label{optLocation}
\begin{align}\vspace{-0.1cm}
\vspace{-0.5cm}
x_i^* = \frac{\int \int_{\mathcal{A}_i}x~ Z_i(x,y)~  \dif x \dif y  }{\int \int_{\mathcal{A}_i} Z_i(x,y)~ \dif x \dif y }, \\ 
y_i^* = \frac{\int \int_{\mathcal{A}_i}y~ Z_i(x,y)~  \dif x \dif y  }{\int \int_{\mathcal{A}_i} Z_i(x,y)~ \dif x \dif y }.
\end{align}
\end{subequations}
Although the objective function $P_i^c +  P_i^t $ is convex with respect to $x_i$ and $y_i$, 
deriving a closed-form solution of (\ref{powerMinEachTwo}), which minimizes both the transmit and mobility power for each UAV, is challenging. 
However, it is easy to find the optimal solution of (\ref{powerMinEachTwo}) based on a gradient-based algorithm.  
Using a similar approach, we can find the optimal location for scenario two.

\section{Simulation Results and Analysis}\label{simulation}

For simulations, we consider a UAV cellular network operating in a  $5$ GHz frequency band for downlink communications. The total available bandwidth for each UAV is $10$ MHz.  
The noise power spectral is set to $-174 $ dBm/Hz. 
For each UAV, the antenna gain is $10$ dB, and the rate  of energy consumption for moving per meter is $\gamma=0.1$ Joules per meter.     
For ML, we use $\frac{7}{8}$ of the dataset to train the model, and the remaining  $\frac{1}{8}$ data is used to  evaluate the performance.

\begin{figure}[!t]
	\begin{center} 
		\includegraphics[width=8.5cm]{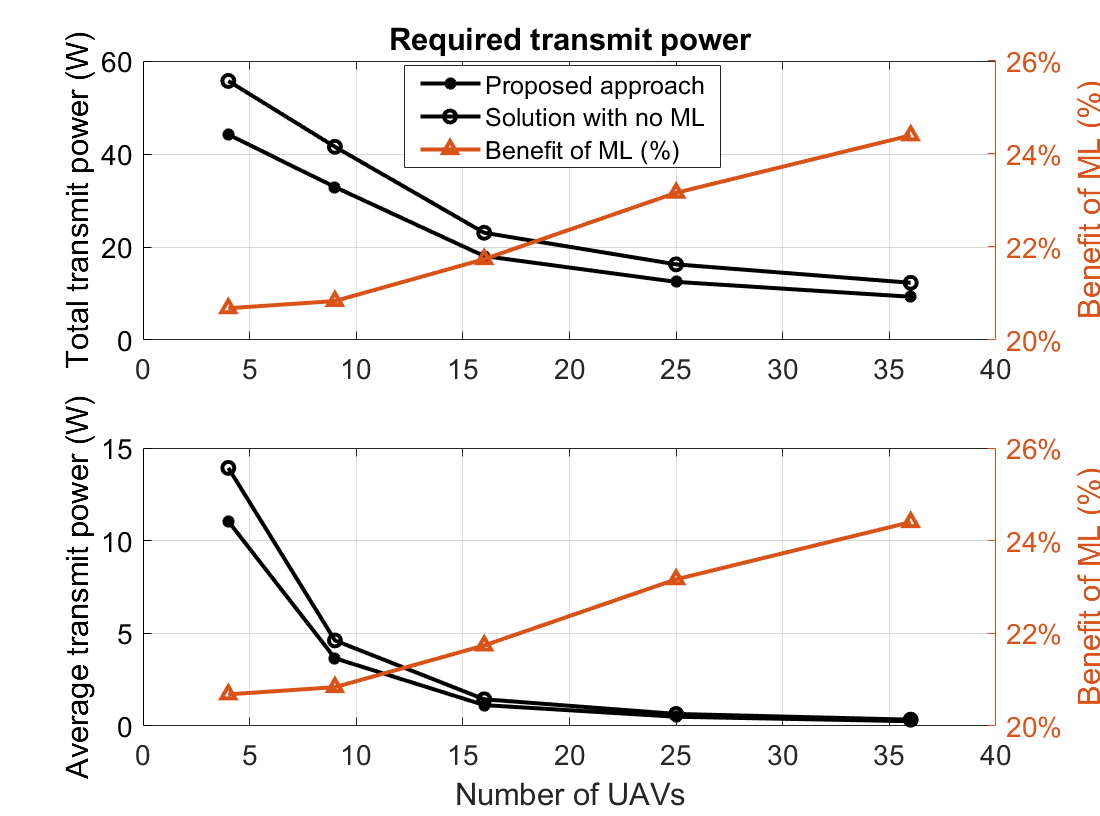} 
		\caption{\label{power}\small Total and average required power for data transmission.}
	\end{center}\vspace{-0.3cm}
\end{figure}
Fig. \ref{power}  shows the total and average communication power per UAV  required to satisfy the users' data demands for two scenarios: the proposed approach and a solution with no ML predictions.
In each case, the proposed optimal partition of service areas and the optimal location deployment are employed.  
Fig. 1  shows that, as the number of UAVs increases, both the total required power and the average communication power will decrease. 
When more UAVs are available, each aerial BS can serve a smaller coverage area, yielding a lower average path loss. Therefore, the needed total transmit power decreases, given a fixed amount of the total cellular traffic.  
As the total required transmit power decreases, the average power reduces accordingly. 
Fig. 1 further shows that compared with the solution without ML predictions, the proposed approach yields a significant improvement of power consumptions. The power reduction varies from $20.68\%$  to $24.40\%$, as the number of UAVs increases from $9$ to $36$. 

\begin{figure}[!t]
	\begin{center} 
		\includegraphics[width=8cm]{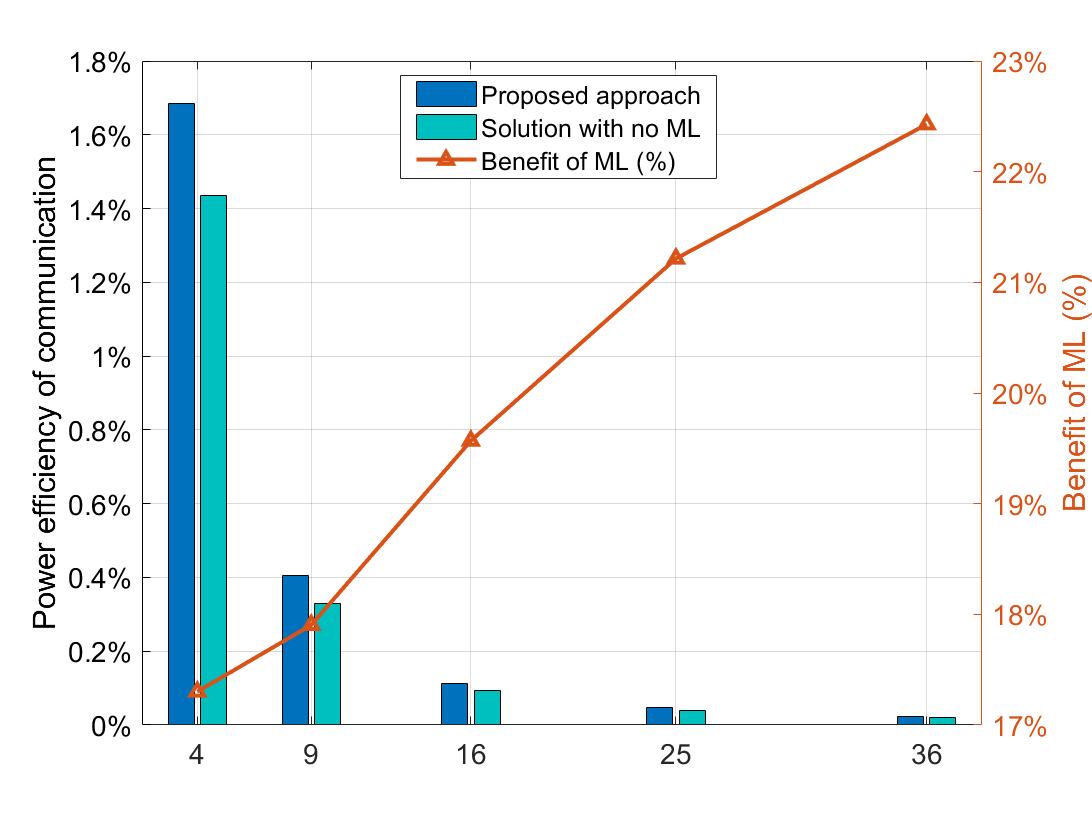}
		\vspace{-0.4cm}
		\caption{\label{energyEffi}\small Average power efficiency of UAV wireless communications over total power consumption. }
	\end{center}\vspace{-0.6cm}
\end{figure}

Fig. \ref{energyEffi} shows the power efficiency, defined as the average percentage of the  transmit power $P_c$ out of total power $P_c+P_t$. 
As the number of UAVs increases, the power efficiencies in both scenarios will decrease. 
Here, we note that UAV mobility will often require more power than wireless transmission. 
By deploying more UAVs, network operator is more likely to send a UAV to meet an instant communication in  a relatively far  hotspot area, which causes more power consumed for mobility. 
Also, as shown in Fig. \ref{power}, more UAVs requires using a less  communication power $P_c$, which further reduces the power efficiency.  
Moreover, Fig. \ref{energyEffi} shows that compared with the solution without ML, the proposed method can improve the power efficiency of UAV communication by up to $22.34\%$.  

\begin{figure}[!t]
	\begin{center} 
		\includegraphics[width=8cm]{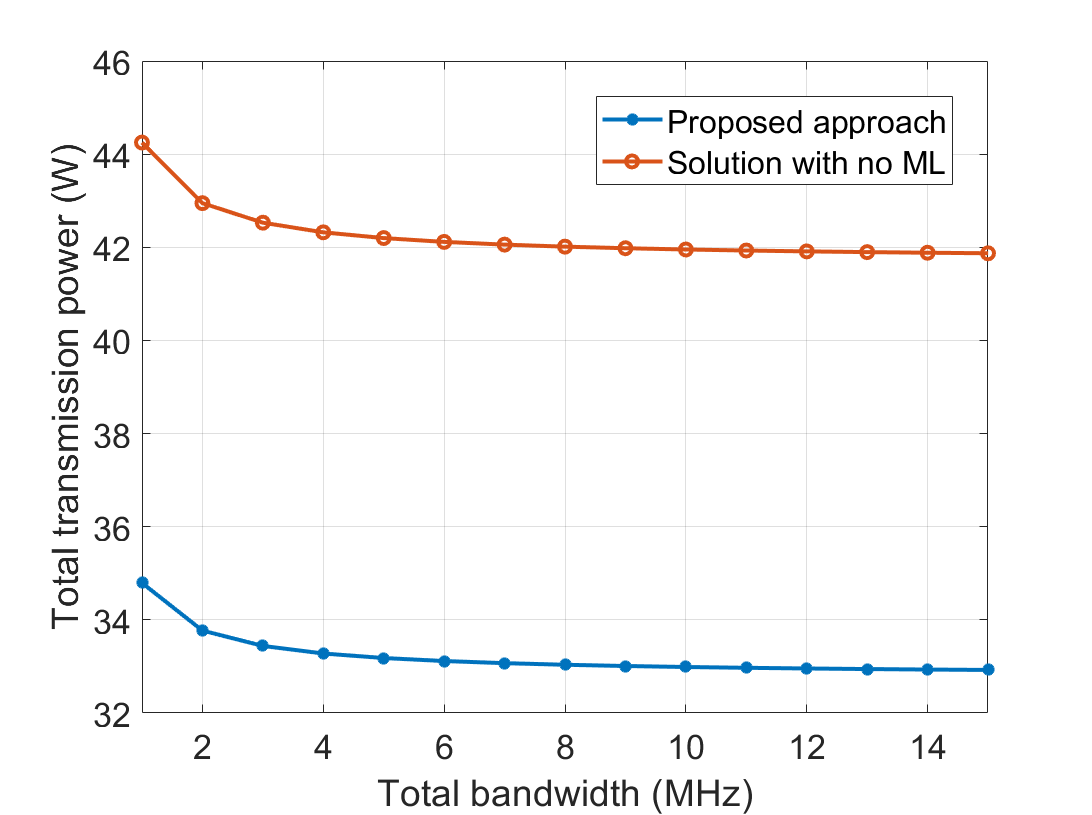}
		\vspace{-0.2cm}
		\caption{\label{spectrum}\small Required transmit power as a function of the total bandwidth.}
	\end{center}\vspace{-0.5cm}
\end{figure}

Fig. \ref{spectrum}  shows the required transmit power as a function of the total bandwidth, assuming nine UAVs.   
As the available bandwidth increases, the transmit power will decrease. However, a wider bandwidth results in a higher noise power, which prevents the reduction of transmit power, especially when the bandwidth is greater than $5$ MHz.  
For such noise-sensitive system, a lower spectrum efficiency cannot save additional power. 
 
\section{Conclusion} \label{conclusion}  
In this paper, we have proposed a novel approach for predictive deployment of UAV aerial BSs to provide an on-demand wireless service to the cellular users. 
We have formulated a power minimization problem to optimize the partition of the service area of each UAV,   while minimizing the UAV power needed for downlink communications and mobility. 
In order to predict hotspots, a novel ML framework based on GMM and WEM has been developed. 
The results have shown that the proposed ML approach can reduce the required downlink transmit power, and improve the average power efficiency by over $20\%$, compared with an optimal deployment of UAVs with no ML prediction.

\def\baselinestretch{0.8}
\bibliographystyle{IEEEtran}
\bibliography{ms}

\begin{thebibliography}{10}
\providecommand{\url}[1]{#1}
\csname url@samestyle\endcsname
\providecommand{\newblock}{\relax}
\providecommand{\bibinfo}[2]{#2}
\providecommand{\BIBentrySTDinterwordspacing}{\spaceskip=0pt\relax}
\providecommand{\BIBentryALTinterwordstretchfactor}{4}
\providecommand{\BIBentryALTinterwordspacing}{\spaceskip=\fontdimen2\font plus
\BIBentryALTinterwordstretchfactor\fontdimen3\font minus
  \fontdimen4\font\relax}
\providecommand{\BIBforeignlanguage}[2]{{%
\expandafter\ifx\csname l@#1\endcsname\relax
\typeout{** WARNING: IEEEtran.bst: No hyphenation pattern has been}%
\typeout{** loaded for the language `#1'. Using the pattern for}%
\typeout{** the default language instead.}%
\else
\language=\csname l@#1\endcsname
\fi
#2}}
\providecommand{\BIBdecl}{\relax}
\BIBdecl

\bibitem{rangan2014millimeter}
S.~Rangan, T.~S. Rappaport, and E.~Erkip, ``Millimeter-wave cellular wireless
  networks: Potentials and challenges,'' \emph{Proceedings of the IEEE}, vol.
  102, no.~3, pp. 366--385, Feb 2014.

\bibitem{bhushan2014network}
N.~Bhushan, J.~Li, D.~Malladi, R.~Gilmore, D.~Brenner, A.~Damnjanovic,
  R.~Sukhavasi, C.~Patel, and S.~Geirhofer, ``Network densification: the
  dominant theme for wireless evolution into 5{G},'' \emph{IEEE Communications
  Magazine}, vol.~52, no.~2, pp. 82--89, Feb 2014.

\bibitem{mozaffari2018tutorial}
M.~Mozaffari, W.~Saad, M.~Bennis, Y.-H. Nam, and M.~Debbah, ``A tutorial on
  {UAV}s for wireless networks: Applications, challenges, and open problems,''
  \emph{arXiv preprint arXiv:1803.00680}, 2018.

\bibitem{mozaffari2016optimal}
M.~Mozaffari, W.~Saad, M.~Bennis, and M.~Debbah, ``Optimal transport theory for
  power-efficient deployment of unmanned aerial vehicles,'' in \emph{Proc. of
  IEEE International Conference on Communications (ICC)}, Kuala Lumpur,
  Malaysia, May 2016, pp. 1--6.

\bibitem{chen2017machine}
M.~Chen, U.~Challita, W.~Saad, C.~Yin, and M.~Debbah, ``Machine learning for
  wireless networks with artificial intelligence: A tutorial on neural
  networks,'' \emph{arXiv preprint arXiv:1710.02913}, 2017.

\bibitem{mozaffari2016efficient}
M.~Mozaffari, W.~Saad, M.~Bennis, and M.~Debbah, ``Efficient deployment of
  multiple unmanned aerial vehicles for optimal wireless coverage,'' \emph{IEEE
  Communications Letters}, vol.~20, no.~8, pp. 1647--1650, Jun 2016.

\bibitem{becvar2017performance}
Z.~Becvar, M.~Vondra, P.~Mach, J.~Plachy, and D.~Gesbert, ``Performance of
  mobile networks with {UAV}s: Can flying base stations substitute ultra-dense
  small cells?'' in \emph{Proc. of 23th European Wireless Conference}, Dresden,
  Germany, May 2017, pp. 1--7.

\bibitem{li2016energy}
K.~Li, W.~Ni, X.~Wang, R.~P. Liu, S.~S. Kanhere, and S.~Jha, ``Energy-efficient
  cooperative relaying for unmanned aerial vehicles,'' \emph{IEEE Transactions
  on Mobile Computing}, vol.~15, no.~6, pp. 1377--1386, Aug 2016.

\bibitem{sharma2016uav}
V.~Sharma, M.~Bennis, and R.~Kumar, ``{UAV}-assisted heterogeneous networks for
  capacity enhancement,'' \emph{IEEE Communications Letters}, vol.~20, no.~6,
  pp. 1207--1210, 2016.

\bibitem{zeng2017energy}
Y.~Zeng and R.~Zhang, ``Energy-efficient {UAV} communication with trajectory
  optimization,'' \emph{IEEE Transactions on Wireless Communications}, vol.~16,
  no.~6, pp. 3747--3760, Mar 2017.

\bibitem{horn2012neural}
J.~F. Horn, E.~M. Schmidt, B.~R. Geiger, and M.~P. DeAngelo, ``Neural
  network-based trajectory optimization for unmanned aerial vehicles,''
  \emph{Journal of Guidance, Control, and Dynamics}, vol.~35, no.~2, pp.
  548--562, Mar-Apr 2012.

\bibitem{chen2017learning}
J.~Chen, U.~Yatnalli, and D.~Gesbert, ``Learning radio maps for {UAV}-aided
  wireless networks: A segmented regression approach,'' in \emph{Proc. of IEEE
  International Conference on Communications (ICC)}, Paris, France, July 2017,
  pp. 1--6.

\bibitem{zhang2017traffic}
S.~Zhang, S.~Zhao, M.~Yuan, J.~Zeng, J.~Yao, M.~R. Lyu, and I.~King, ``Traffic
  prediction based power saving in cellular networks: A machine learning
  method,'' in \emph{Proc. of the 25th ACM SIGSPATIAL International Conference
  on Advances in Geographic Information Systems}, Redondo Beach, CA, USA, Nov
  2017, p.~29.

\bibitem{gebru2016algorithms}
I.~D. Gebru, X.~Alameda-Pineda, F.~Forbes, and R.~Horaud, ``{EM} algorithms for
  weighted-data clustering with application to audio-visual scene analysis,''
  \emph{IEEE Transactions on Pattern Analysis and Machine Intelligence},
  vol.~38, no.~12, pp. 2402--2415, Jan 2016.

\bibitem{al2014optimal}
A.~Al-Hourani, S.~Kandeepan, and S.~Lardner, ``Optimal {LAP} altitude for
  maximum coverage,'' \emph{IEEE Wireless Communications Letters}, vol.~3,
  no.~6, pp. 569--572, July 2014.

\bibitem{cityCellularTrafficMap}
``City cellular traffic map,''
  \url{https://github.com/caesar0301/city-cellular-traffic-map}, accessed:
  2016-10-05.

\bibitem{paul2011understanding}
U.~Paul, A.~P. Subramanian, M.~M. Buddhikot, and S.~R. Das, ``Understanding
  traffic dynamics in cellular data networks,'' in \emph{Proc. of IEEE
  International Conference on Computer Communications (INFOCOM)}, Shanghai,
  China, Apr 2011, pp. 882--890.

\bibitem{christopher2016pattern}
M.~B. Christopher, \emph{Pattern recognition and machine learning}.\hskip 1em
  plus 0.5em minus 0.4em\relax Springer-Verlag New York, 2016.

\bibitem{mozaffari2017wireless}
M.~Mozaffari, W.~Saad, M.~Bennis, and M.~Debbah, ``Wireless communication using
  unmanned aerial vehicles ({UAV}s): Optimal transport theory for hover time
  optimization,'' \emph{IEEE Transactions on Wireless Communications}, vol.~16,
  no.~12, pp. 8052--8066, Apr 2017.

\end{thebibliography}

\end{document}